\documentclass{aa}

\begin{document}
\input psfig.tex
 
\title{Dim galaxies and outer halos of galaxies missed by 2MASS ?
The near--infrared luminosity function and density}

\author{S. Andreon}
\institute{Osservatorio Astronomico di Capodimonte, Naples, Italy }


\titlerunning{Flux lost by 2MASS}

\date{A\&A in press}
 
\abstract{By using high--resolution and deep $K_s$ band observations of
early--type galaxies of the nearby Universe and of a cluster at $z=0.3$ we show
that the two luminosity functions (LFs) of the local universe derived from
2MASS data miss a fair fraction of the flux of the galaxies (more than 20 to 30
\%) and a whole population of galaxies of central brightness fainter than the
isophote used for detection, but bright enough to be included in the published
LFs. In particular, the fraction of lost flux increases as the galaxy
surface brightness become fainter.
Therefore, the so far derived LF slopes and characteristic luminosity as
well as luminosity density are underestimated. Other published near--infrared
LFs miss flux in general, including the LF of the distant field computed in a
3 arcsec aperture.
 } \maketitle 

\section{Introduction}

The luminosity function (LF) is the benchmark against which theories of galaxy
formation and evolution in a variety of cosmological models can be tested.
Therefore, the LF is fundamental to observational cosmology and theory of
galaxy formation. In particular, the near infrared LF is a good tracer of
evolved stellar populations and hence of the total stellar content of
galaxies,  much better than optical LFs affected by dust extinction and young
stellar populations. Near--infrared luminosities are more directly related to
stellar mass, constraining both the history of the star formation and the
galaxy formation models (see, e.g., Cole et al. 2000 and references therein)

The luminosity density, which is the integral of the luminosity weighted by the
LF, is an important input to estimates of star formation history of the
universe, its chemical evolution and of the extragalactic background ligh.

The recent release of near--infrared imaging data of a large fraction of the
sky by 2MASS prompt two groups to derive the near--infrared of the local
universe: Kochanek et al. (2001), by using literature or new redshift data, 
compute the LF of a very nearby sample of galaxies spread over a large sky
area, while Cole et al. (2001) coupled near--infrared data to the 2dF redshift
survey and studied a deeper sample over a smaller area.

Kochanek et al. (2001) LF has been computed by adopting isophotal magnitudes at
the 20.0 mag arcsec$^{-2}$, $K_{20}$, of a large sample ($\sim4000$) of
galaxies selected to have $K_{20}<11.25$ mag, extracted from the 2MASS extended
object catalog (Jarrett et al. 2000). A flat LF ($\alpha\sim-0.9$) has been
found both for early--type galaxies and for late--type galaxies separately, and
($\alpha\sim-1$) for the whole sample.  Cole et al. (2001) LF also uses 2MASS
data and found similar results in terms of slope ($\alpha=-1.0$). They used a
Kron--like $J$ magnitude and $J-K$ aperture color for measuring the $K$ flux
of objects. Both works based on 2MASS photometry found flat slopes.

The found slope is significatively shallower than the one derived by Andreon \&
Pell\'o (2000) for the Coma cluster LF in the $H$ band ($\alpha\sim-1.3$),  and
from that derived for the cluster AC\,118 (Abell 2744) at $z=0.3$ in the $K_s$
band ($\alpha\sim-1.3$), for a sample of galaxies outside the cluster core
where cluster--related effect should be low.  


Furthermore, Wright (2001) notes that the luminosity density based on 2MASS data
are between 1.4 and 2.5 times fainter than the one expected by assuming the SLOAN
luminosity densities (derived from Blanton et al. 2001 LFs) and a typical
spiral spectrum. A redder, elliptical--like or dusty, spectrum would only
increase this disagreement.

In this paper we check whether the flat slope and the low luminosity density
derived from 2MASS data are affected by flux (and galaxy) losses  because of
the shallower data set used (3 sec exposures). We also show that other LFs
computed by using aperture magnitudes are skewed with respect to the true one.

In this paper we assume $H_0=50$ km s$^{-1}$ Mpc$^{-1}$ and $q_0=0.5$, but
the choice of the cosmology is largely irrelevant.

\section{LF dependence on the choice of the magnitude adopted}

\subsection{The data}

In order to understand if a significant fraction of the galaxy flux is missed
in the near--infrared,  we consider two comparison datasets: Pahre (1999)
galaxies and AC\,118 galaxies. Pahre' galaxies are in the nearby universe, are
normal and common early--type galaxies obeying to the Fundamental Plane and
have been observed by pointed observations deeper and of higher resolution 
than the
2MASS survey data. Pahre (1999) lists effective radius $r_e$ and brightness
$\mu_e$\footnote{Pahre (1999) lists mean surface brightness within $r_e$, which
is equal to $\mu_e-1.39$ adopting a de Vaucouleurs (1948) law.} and total
magnitude for his galaxies. All types of magnitude (isophotal, aperture, Kron,
etc.) can be easily computed, assuming that the galaxy profile is accurately
described by a de Vaucouleurs (1948) law. This assumption is the major
limitation of this dataset.

The second dataset is based on high resolution and deep near--infrared images
of the cluster AC\,118 (Andreon 2001) at $z\sim0.3$. This dataset is a well
controlled one being a volume complete sample, but the rest--frame spatial 
resolution is much worser than for the Pahre' sample because of the large
cluster distance. Unlike the Pahre' sample, this dataset includes galaxies of
all morphological types, even if early--type galaxies are the majority, being a
cluster sample. For the images of these galaxies we directly  compute several
types of magnitudes (isophotal, aperture and Kron--like) from pixel values,
thus avoiding the principal assumption done adopting the Pahre' sample.
In this respect the two samples are complementary.

\subsection{2MASS isophotal magnitudes and the Kochanek et al. LF}

Most of the Pahre' galaxies are in the same Universe volume studied by Kochanek
et al. (2001). With respect the
Kochanek's et al. sample, the Pahre' sample is 1.5 mag deeper, but we found
that when culled at the same apparent magnitude shows the same absolute
magnitude distribution. Therefore, the Pahre' sample does not grossly
undersample galaxies of any absolute magnitude present in the Kochanek's et al.
sample although it is not a well controlled sample, as the latter. By numerally
integrating the galaxy profile down to the 20 mag arcsec$^{-2}$ isophote we can
compute the fraction of the detected flux by 2MASS data. Figure 1 shows the
fraction of flux within the 20 mag arcsec$^{-2}$ isophote as a function of the
object magnitude. The missed flux could be as large as 70 \%.  For about 60 \%
of the sample it is larger than 30 \% and larger than 15 \% for more than 97 \% of
the sample, much larger that the value claimed by Kochanek et al. (10 to 20 \%
for most of the galaxies). Figure 2 shows that the fraction of flux lost is
mainly a function alone of the galaxy surface brightness. This holds because
the dependence on the effective radius of the fraction of missed flux can be
factorized and reduced in the plotted ratio. Galaxies, even bright ($K\sim10$)
ones, have effective brightness not too much different from the 20 mag
arcsec$^{-2}$ detection isophote and a significant part of their flux is lost
below the detection isophote. In particular, 50 \% of the flux is lost, by
definition, when $\mu_e=20$ mag arcsec$^{-2}$, i.e. when the detection and
effective isophotes are the equal. 

The absolute magnitude dependence of the missed flux is negligible for the
Pahre' sample. However, this dependence cannot be  definitively ruled out
because the Pahre' samples is not complete and the dependence is expected via
the correlation between absolute magnitude and surface brightness of galaxies
(faint galaxies tend to be of low surface brightness, Andreon \& Cuillandre
2001).

To summarize, the Pahre' sample shows that 2MASS isophotal mag lost a
significant part of the galaxy flux, larger than the claimed 10 to 20 \%. 
Therefore, characteristic luminosity, the luminosity density  and possibly the
slope of the LF are underestimated by adopting this isophotal magnitudes.

Cole et al. (2001) independently confirm that the 2MASS isophotal mag missed a
fair fraction of the galaxy flux. They show that the 2MASS isophotal mag misses
some 0.1 mag with respect the Kron--like mag listed in the 2MASS database,
which in turn misses about 0.15 mag with respect the true Kron mag (because
computed on a too small object region), which in turn misses 5 to 10 \% of the
total flux. Therefore the flux lost by the 2MASS isophotal magnitude is 0.3 to
0.35 mag or, 25 \% to 30 \% of the galaxy flux, in reasonable agreement with
our estimate.

Let us now consider the second dataset.
Figure 3 compares our ``total" $K_s$ magnitude (that will be defined in the
next section) vs the isophotal magnitude within the 21.5 mag arcsec$^{-2}$,
that correspond to 20.0 mag arcsec$^{-2}$ isophote in the rest--frame, when 
cosmological dimming and k--correction are taken into account. The dotted line
in Fig 3 is the bisector line. The rest--frame 20.0 mag arcsec$^{-2}$ isophotal
magnitude is always fainter than our ``total" magnitude, by 0.5 mag on average,
which in turn is, of course, fainter than the true total magnitude. Therefore,
this plot shows that the isophotal magnitude at the 20.0 mag arcsec$^{-2}$
misses some 40 \% flux from the galaxies, in agreement with the previous
estimate based on Pahre' data. A second effect could be appreciated from
Figure 3 by noting the dearth of galaxies at faint magnitudes. There is almost
no galaxy as faint as $M_{K_s}=-22$ mag while the cluster LF is flat (at worst,
see Andreon 2001) at these magnitudes (and the background should also
contribute with some galaxies). This is a re-state of the low surface
brightness problem: when the central surface brightness drop below the
detection isophote the object is undetected. This type of galaxies are bright
enough to be included in the local near--infrared LF but are missed by 2MASS 
because their surface brightness is too dim. Therefore, for galaxies in our
sample, i.e. for an essentially volume complete sample of galaxies in an
intermediate redshift cluster, the rest--frame 20 mag arcsec$^{-2}$ isophote is
not a good choice for measuring the LF for two reasons: because of the large
fraction of missed flux, and because of the numerous missed galaxies at whole.
Both effects produce flat (and faint) LFs and low luminosity densities.

Figure 4 shows that our ``total" magnitude is not bad: for most of the Pahre'
(1999) sample the magnitude within 2.5 Kron radii (Kron 1980) misses an
approximatively constant 10 \% of the galaxy flux. In this specific calculation
we take into account that the flux used to compute the second moment of the
light distribution (the Kron $r_1$ radius) is actually integrated over $12^2$
times (Bertin, 2001, private communication) the detection area by SEx (Bertin
\& Arnout 1996)\footnote{but only in the detection area for 2MASS objects (Jarrett
2001, private communication cited in Cole et al. 2001.)}. For faint objects
($M_K>\sim-22$ mag) we adopted an aperture magnitude, but the aperture is 
large enough to include most of the flux. The ultimate reason for our ``total"
mag being a successful measure of the galaxy total flux is that the Kron radius
is extremely well correlated to the effective radius and
the ratio of the two radii is independent on $\mu_e$ and almost constant.
Therefore, the galaxy flux is integrated within an almost fixed number of
effective radii, which contains an almost constant fraction of the total flux,
for a fixed surface brightness profile shape. Outlyer points in Fig. 4 turn
out to be galaxies with a so large effective radius that the computed Kron
radius is underestimated from the object portion considered by the detection
software, much like the usual situation for 2MASS Kron--like magnitudes.


\subsection{2MASS Kron--like magnitudes and Cole et al. LF}

By using the Pahre' data we can repeat the same previous exercises  for 2MASS
Kron--like magnitudes used for computing the 2dF LF (Cole et al. 2001).  These mags
are measured within 2.5 times the first moment of the light distribution of the
pixels brighter than the detection threshold in J (21.7 mag, Jarrett et al. 1999,
cited in Cole et al. 2001\footnote{ See also 
http://spider.ipac.caltech.edu/staff/jarrett/2mass/repeats/kron.html}) minus the $J-K$ color (computed on a smaller galaxy
area). This particular choice allows  the 2dF team to integrate a larger fraction of
the galaxy flux (than adopting the Kron $K$ mag), under the hypothesis of minor
color gradients between the $J$ and $K$ bands.  Again, mimicking the integration of
the galaxy flux and assuming the observed average color $J-K=1.1$ mag (Cole et al.
2001), we can easily compute the fraction of the flux missed by the magnitudes
adopted by the 2dF team (Figure 5). Unlike SEx, 2MASS computes the moments of the
light distribution on a often tiny fraction of the galaxy profile, giving an
under--estimation of them and, by consequence, of  the 2MASS Kron--like magnitude.
Due to the under--estimation of the true Kron radius (even if measured in the deeper
J band), the flux of galaxies are underestimated by 20 to 50 \%, and 0.35 mag on
average for the Pahre' sample\footnote{A revision of the Kron photometry
is planned for the final reprocessing of the 2MASS data, and therefore
our criticism likely concern exclusively Kron magnitudes in the 2MASS incremental
releases.}.  This result is a bit larger than the value quoted by
Cole et al. (2001): they found that the 2MASS Kron--like mag misses 0.06 mag with
respect to the Loveday (2000) Kron mag (for a sample of common objects), which in
turn misses 10 \% (0.1 mag) of the total flux, as it is claimed by Cole et al.
(2001) and checked by us. Therefore, according to Cole et al. (2001), their mag
misses a total of 0.16 mag (for Pahre' galaxies, vs our estimate of 0.35 mag). 

Judging from their Figure 15, the Cole et al.
(2001) estimate of the lost flux is low: their measured LF is still shallower
and fainter than expected from the SDSS $z^*$ LF (Blanton et al. 2001)
converted in $K$. It is fainter because their correction for missing flux is
underestimated. By correcting the Cole et al. (2001) LF by a further 0.2 mag (our
0.35 mag minus 0.16 mag already corrected for), the SDSS and Cole et al. LFs
are in reasonable agreement at the bright end and the two LFs have very similar
amplitudes ($\phi^*$) at $M^*$. At the faint end, the Cole et al. LF is
shallower because galaxies of low surface brightness are missed because too
dim, and are not listed in the 2MASS catalog.

\subsection{Further checks}

Since some of Pahre' galaxies are listed in the 2MASS database, it is quite easy to
directly measure the fraction of flux lost, because it is given by the difference
between the total magnitude, listed in Pahre' paper as measured from the growth
curve technic, and 2MASS mags listed in the 2MASS catalog. 

Pahre' galaxies are identified by name (about half of them are NGC/IC galaxies), while 2MASS
galaxies by coordinates. From the Pahre' list of 340 galaxies we was able to get sky
coordinates from NED for 327 of them. Then, we look for sources, within a 5 arcsec radius
circle  centered on NED coordinates, in the second 2MASS incremental release catalog. Out of
327 objects, 122 of them (37\%) are listed in the 2MASS catalog.

The left panel of figure 6 shows the fraction of flux lost by the isophotal
magnitude adopted by Kochanek et al., $K_{20}$, as a function of the effective
surface brightness of the galaxy. A
clear trend is present in agreement with what is found in section 2.2: 
the missed flux is larger for lower surface brightness
galaxies, and could be as large as 50 \% when $\mu_e=20$ mag arcsec$^{-2}$, as it
should be when detection and effective isophote are equal. On average,  $K_{20}$
losts 0.2 mag for galaxies listed both in the Pahre' sample and in 2MASS, but, of
course, the average depends on the $\mu_e$ distribution. 

The right panel of  figure 6 shows the fraction of lost flux by the  Kron--like
magnitude adopted by 2dF. The result is quite similar to that found for isophotal
magnitudes, and it is in agreement with what found in section 2.3:  a similar trend
for increasing missing flux when effective brightness become fainter is present,
and, on average 0.2 mag are lost for galaxies listed both in the
Pahre' sample and 2MASS.  

We stress that we are talking about bright, famous and rare galaxies: these galaxies have on
average $K\sim-24.7$ mag. Galaxies fainter by four, or five, magnitudes are included in the
Kochanek et al. and in the 2dF LFs. These ordinary galaxies have fainter surface
brightnesses because of the correlation between absolute magnitude and surface brightness,
as already noted. Therefore, the fraction of missed flux measured by Figure 6 is
underestimated, when an ordinary sample is chosen. 
A rough estimate of the typical fraction of flux lost for these normal galaxies
can be computed as following: the characteristic magnitude is  $K^*\sim-25$ mag
(Kochanek et al. and Cole et al.). These galaxies  turn out to have typical $\mu_e$ of 17
mag arcsec$^{-2}$ (with a very large scatter) in the K band for the Pahre' sample. In the
optical, Sandage \& Perelmuter (1990) shows that galaxies having $M^*+4$ have $\mu_e$ two
mag fainter than $M^*$ galaxies, on average. Assuming negligible color gradient between
optical and infrared colors, galaxies having $M^*+4$ have $\mu_e \sim 19$ mag arcsec$^{-2}$
(with a large scatter) in the $K$ band. The same result can be found assuming a reasonable
color for galaxies, and reading directly the optical $\mu_e$ at $M^*+4$ in, say, Sandage \&
Perelmuter (1990). At such a brightness the fraction of flux lost by isophotal mag is 40 \%
(see Figure 2 or left panel of Figure 6). The estimate for the Kron magnitude adopted by 2dF
is quite the same (see Figure 5 or right panel of Figure 6).

Anyway, Figure 6 definitively shows that the fraction of missed flux is easily much
larger than 15 \%, that it is correlated to $\mu_e$ and that it is already large for
famous high surface brightness galaxies. For ordinary galaxies the fraction of lost
flux is by necessity larger. Results from Figure 6 do not assume a surface
brightness  profile for the  galaxies, while Figure 2 and 5 do, and the agreement
between findings drawn from these figures confirms that the assumption of a de
Vaucouleurs law is a good one.

The galaxy with the largest fraction of lost flux in Figure 6 has $K= 8.4$ mag.

Since Figure 6 is produced without almost any author work (we just paired catalog
entries), the probability that a mistake slip inside this plot is very low.

\medskip  As a final check, we compared our synthetic photometry vs 2MASS measures
for the $K_{20}$ and Kron--like mags. We found an offset of  $<\sim0.1$ mag, with a scatter
of 0.1 mag, in the sense that 2MASS mag are brighter than ours. While the whole
offset is not entirely understood by the author, part of it come from known effects
described in Jarrett et al. (2001, isophotal contours are uncalibrated at 0.1 to 0.2
mag arcsec$^{-2}$), by some likely operation performed on 2MASS images for isophote
regularization (convolution with a kernel) and from the fact of having neglected
seeing effect in our computation. We stress that, even without taking into account
all these effects, the systematic offset is only a bit larger than the photometric
error quoted in the 2MASS catalog (0.08 mag) for Pahre' galaxies.

Our underestimation of the 2MASS flux (or their overestimation with respect our
modelleing) reduces the {\it average} fraction of missed flux lost by 2MASS from about 0.35
mag (claimed in previous sections) to about 0.25 mag for both Kron--like and isophotal
mags. 

The same 0.1 mag offset helps to reduce the disagreement between our and Cole et al. (2001) 
estimate of the fraction of flux lost by their Kron--like magnitude. In fact, when this
offset is take in to account, the two estimates differs by 0.1 mag only.

Nevertheless, Figure 6 unambigoulsly shows that the fraction of flux lost depends
on $\mu_e$, and could be very large even for galaxies bright and famous 
enough to have a name. From more normal galaxies, the same figure shows that
the average flux lost is by necessity large.

\subsection{Aperture magnitude and LFs based on them}

The problem described for the local field LF is, in fact, a general one. For AC\,118
galaxies (at $z=0.3$) our (Andreon 2002) 3 arcsec aperture magnitude (16 Kpc for
galaxies at the AC\,118 redshift) is not a surrogate for ``total magnitude": we
checked that a lot of flux is lost by comparing 3 arcsec aperture magnitude to both
our surrogate of ``total" magnitude (figure not shown) and to total mags of Pahre'
galaxies (redshifted at the cluster redshift, figure not shown). Therefore the LF
computed by using this aperture is skewed with respect to the true one. Of course,
the same holds for the {\it field} LF too. Using again the Pahre' sample we can
compute the fraction of flux missed adopting a 3 arcsec aperture for galaxies at
$z=0.6$ (Figure 7). Even if we recover the know result that, on average, the
fraction of missed flux is 0.2 mag (value for which the mag of field galaxies are
corrected for), this holds only at intermediate absolute magnitude. For bright
galaxies the needed correction is quite large and, most important, the flux lost is
large  because these galaxies are bright. This flux lost possibly represent a fair
fraction of the luminosity density. At the contrary the applied correction is too
large for faint galaxies.

Therefore, the 3 arcsec aperture magnitude, corrected to 6 arcsec aperture by a
single average offset (e.g. Cowie et al. 1996), and other similar
aperture--corrected magnitudes used for computing the field LF at intermediate
redshift are coarse approximate of the total mag.
This approximation holds maybe
for galaxies of the same absolute magnitude (possibly of similar brightnesses
and scales), but not for galaxies of quite different magnitudes which differ in
brightness and scale and lay in different parts of the LF. These approximations
are minor problems when errorbars are large because of the smallness of the
sample, but when the sample is large, as it is the case of the present--day
near--infrared LFs, systematic errors are the largest sources of
uncertainty. 


\section{Discussion}

By using the Pahre (1999) sample of early--type galaxies lying in the
near--universe and obeying to the Fundamental Plane and our high resolution and
deep near--infrared images of galaxies at intermediate redshift we show that 
the two recently determined near--infrared LFs of the nearby Universe based on
2MASS data (Kochanek et al. 2001 and Cole et al. 2001) suffer from flux lost
below the detection isophote and by missing galaxies of low surface brightness,
but bright enough to be included in the LF. 

 Three paths have been followed for computing the fraction of missing flux: 

-- we directly compare published total mag to mag listed in the 2MASS catalog
for NGC/IC galaxies in common between Pahre (1999) and the 2MASS second 
incremental release, 

-- we simulate the 2MASS magnitude measurement
for all the Pahre' galaxies by synthetic
photometry by assuming a de Vaucouleurs law for their surface brightness profile
and we derive the fraction of missed flux, and

-- we use actual $K$ images of galaxies in an intermediate redshift cluster. 

All three paths give the same results: the fraction of missed flux by isophotal or
Kron--like magnitudes is correlated to galaxy effective surface brightness and could
be very large for galaxies included in the Kockanek et al. and 2dF LFs. 
It is larger
than quoted in Kochanek et al. and Cole et al. for galaxies in the Pahre' list,
($\sim0.35$ mag from syntetic photometry over the whole sample, $\sim0.25$ mag
from direct comparison of the galaxies in common with the 2MASS database),
that are famous and bright enough to have a name. For more normal galaxies
the fraction of the flux lost is larger (Figure 2, 5 and 6)
due to the correlation between absolute magnitude and surface brigthness.

The dearth of faint galaxies in Figure 3, coupled with a flat (at worst)  AC\,118 LF
shows that galaxies of low surface brightness, but bright enough to be included in
the Kochanek et al. and Cole et al LFs, exist but are undetected by 2MASS. This is
also the extreme consequence of the previous pointed effect, when the missed flux is
equal to the total flux. Therefore, the slope of the two mentioned LFs is
underestimated.

Furthermore, a significant loss of flux has been shown in this paper for the
near--infrared LFs based on 3" aperture mag, as usually adopted for LFs at
intermediate redshifts (and also for computing galaxy counts).

The missed flux has, of course, an obvious relevance to the determination of the
luminosity density, which, beside the systematic errors shown in this paper,
have presently insignificant statistical errors (Cole et al. 2001). Some
cosmological consequence of missed flux is reported in Wright (2001).

In general, our consideration on the relevance of the type of magnitude used
for the LF computation are confirmed in the optical window by Blanton et al.
(2001), that show how much the so far derived optical LFs, such as the Las
Campanas (Lin et al. 1996) and 2dFGRS (Folkes et al. 1999), are skewed with
respect the true ones because of the use of isophotal magnitudes. Another
similar finding is reported in Garilli, Maccagni \& Andreon (1999): the slope
of the LF for a sample of 2200 galaxies is shallower when a 20 Kpc aperture
magnitude is used in place of pseudo--total magnitudes.

Summarizing, when the sample is large, of the order of several hundred galaxies,
the largest errors on the LF are systematic in nature. A similar conclusion has
been suggested by Kochanek, Pahre \& Falco (2001) for some LFs splitted by
spectral types (ESP, Zucca et al. 1997; Las Campanas, Lin et al. 1996, Bromley
et al. 1998; 2dFGRS, Folkes et al. 1999).

Can the true LF be recovered from the skew one?

First of all, the faint end can be hardly recovered because  galaxies missed by
2MASS are not listed at all, and therefore it is unknow how many of them are
missing. By using a deeper sample  it is possible to determine the absolute
magnitude at which the population of galaxies missed by 2MASS become important and
one can limit the LF determination to brighter magnitudes. The completness can
be determined, say,  by comparing standard and stacked 2MASS observations   (for
example the cluster Abell 3358 has been scanned 30 times by 2MASS), or the 2MASS catalog to
published {\it complete} deeper near--infrared catalogs (such as Andreon et al. 2000
for the Coma cluster). Our present analysis allow to measure how much flux is lost
per galaxy (detected or not), but not how many galaxies are lost.

At brighter magnitudes, the ``total" magnitude could be
recovered by using a measure of the galaxy scale, but with some approximation.
In fact,  the correction from measured mag to total mag depends on the galaxy
growth curve, that in turn depends on the scale and brightness of each individual galaxy
even under the simplistic assumption that the surface brightness profile of
all galaxies is well described by a universal law. Therefore, at least a
measure of the galaxy scale is needed for estimating the LF slope and
characteristic luminosity. Such a measure is encoded in the 2MASS data 
products, although it is biased low because the moments of the ligh
distribution
are computed over often a tiny part of the galaxy surface brightness profile.
The final processing of the 2MASS data will use a more elaborate
schema for computing Kron radii and magnitudes, as described in 
http://spider.ipac.caltech.edu/staff/jarrett/2mass/repeats/kron.html~.

A rough estimate of how much the LF is skewed by adopting magnitudes that are
know to miss flux (and whole galaxies), can be guessed by comparing two
determinations of the LF of AC\,118: the first one is computed by using
a 3 arcsec aperture (Andreon 2002) and agrees with 
both determinations of the local LF, while
the second one adopt our surrogate of ''total'' mag (Andreon 2001): the LF is
shallower by 0.3 in $\alpha$ and fainter by 0.9 mag in $M^*$ (but be aware that
errors on best fit parameters are strongly coupled and therefore  other pairs
of values give almost equivalent descriptions of the difference).

\begin{acknowledgements} This work is dedicated to my daughters, Chiara and
Marta Andreon, the new twin stars born under the sign of Gemini.  Discussion with
A. Wolter, R. Rampazzo and A. Iovino is acknowledged. This work has been performed
at the Osservatorio Astronomico di Brera that I thank for hospitality. The
author acknowledges the referee, for suggesting the comparison
presented in section 2.4..

This publication makes use of data products from the Two Micron All Sky Survey,
which is a joint project of the University of Massachusetts and the Infrared
Processing and Analysis Center/California Institute of Technology, funded by the
National Aeronautics and Space Administration and the National Science Foundation
This research has made use of the NASA/IPAC Extragalactic Database (NED) which is
operated by the Jet Propulsion Laboratory, California Institute of Technology, under
contract with the National Aeronautics and Space Administration. 

\end{acknowledgements}


\begin{figure*}
\psfig{figure=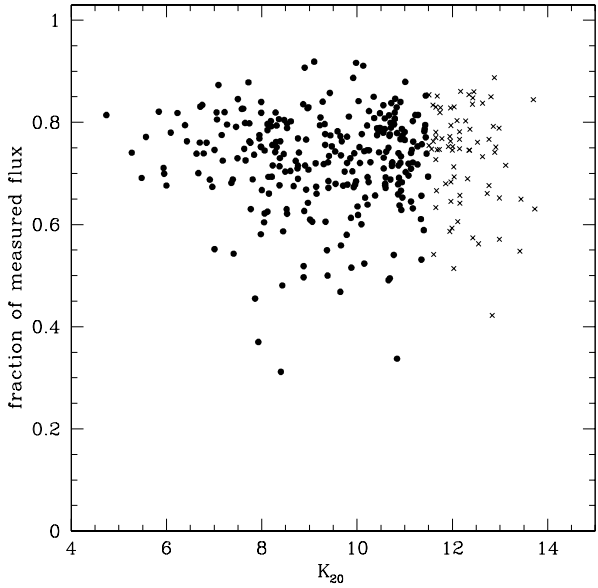,height=8truecm}
\caption[h]{
Fraction of the flux inside the 
20 mag arcsec$^{-2}$ isophote vs apparent magnitudes for
galaxies in Pahre (1999). Filled points are for galaxies with 
$K<11.5$ mag, which mimics the
Kochanek et al. $K_{20}<11.25$ mag selection (the slightly difference in
limiting mag take into account the fact that the isophotal mag miss some flux).
Crosses are for fainter galaxies.
The isophotal magnitude losts at least 20 \% of the flux. 
This and the following plots (with the exceptions
of Figure 3 and 6) consider about 300 galaxies.
}
\end{figure*}

\begin{figure*}
\psfig{figure=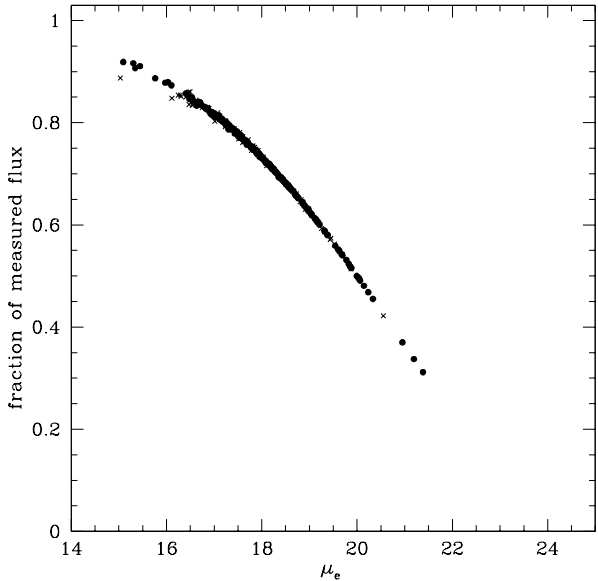,height=8truecm}%
\caption[h]{
Fraction of the flux inside the 20 mag arcsec$^{-2}$ isophote
vs effective surface brightness for
galaxies in Pahre (1999). Symbols are as
in the previous figure. The reason
why the flux is lost is evident: even bright (see previous Figure) 
galaxies have $\mu_e$
few mag brighter than the threshold at which the flux is integrated 
(20 mag arcsec$^{-2}$) and therefore the profile is integrated over a small surface
brightness range.}
\end{figure*}

\begin{figure*}
\psfig{figure=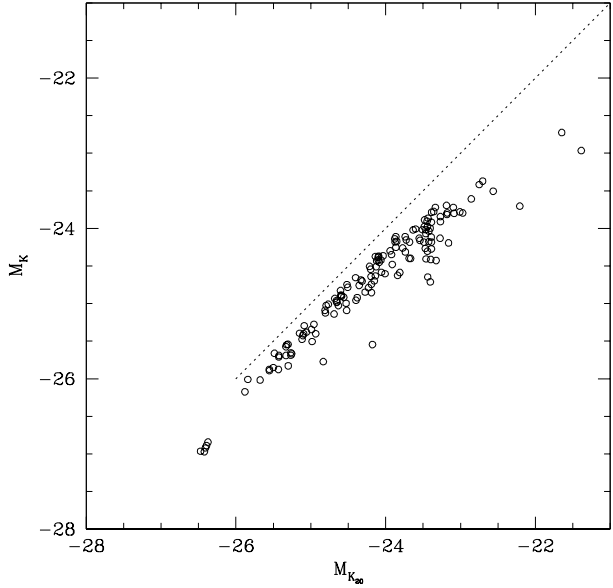,height=8truecm}
\caption[h]{
Our ``total" vs isophotal ($\mu=21.5$ mag arcsec$^{-2}$)
magnitudes of galaxies in the AC\,118 direction. The adopted isophote
correspond to the rest--frame 20 mag arcsec$^{-2}$ isophote for
galaxies at the AC\,118 distance. The dotted
line is the one--to--one relation. The isophotal mag misses a significant fraction of the galaxy flux, 
because
isophotal mag are much fainter than ''total" mag. 
Note the dearth of faint galaxies (as opposed to the flat of slightly rising
cluster LF): their central brightness is so low that they are not detected at all.
Therefore, isophotal magnitude misses also whole galaxies.}
\end{figure*}

\begin{figure*}
\psfig{figure=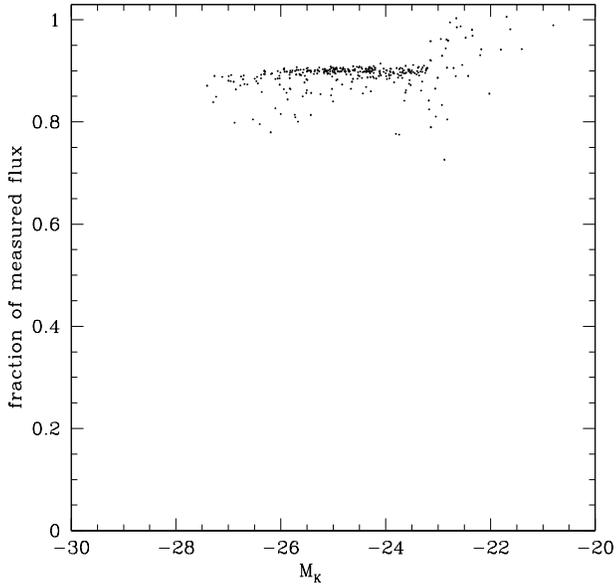,height=8truecm}%
\caption[h]{
Fraction of the flux within our ``total" magnitude vs absolute magnitude for
the Pahre' (1999) sample.}
\end{figure*}

\begin{figure*}
\psfig{figure=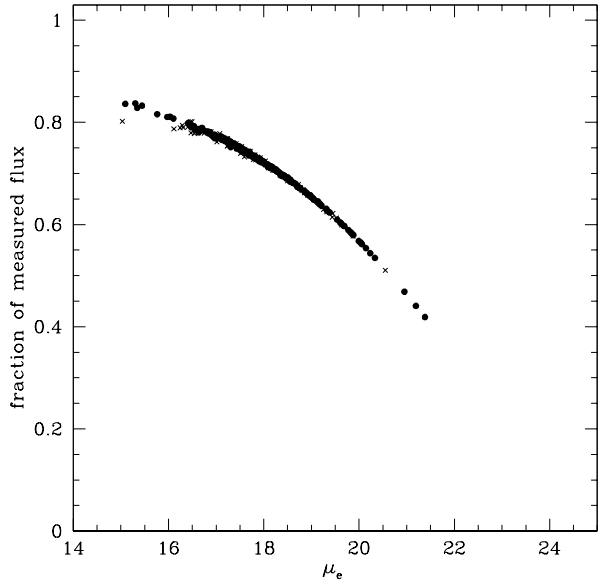,height=8truecm}%
\caption[h]{
Fraction of the flux inside the K mag adopted for the 2dF
near--infrared luminosity. Note the similarity with respect Fig. 2, except
for a milder surface brightness dependence.}
\end{figure*}

\begin{figure*}
\hbox{\psfig{figure=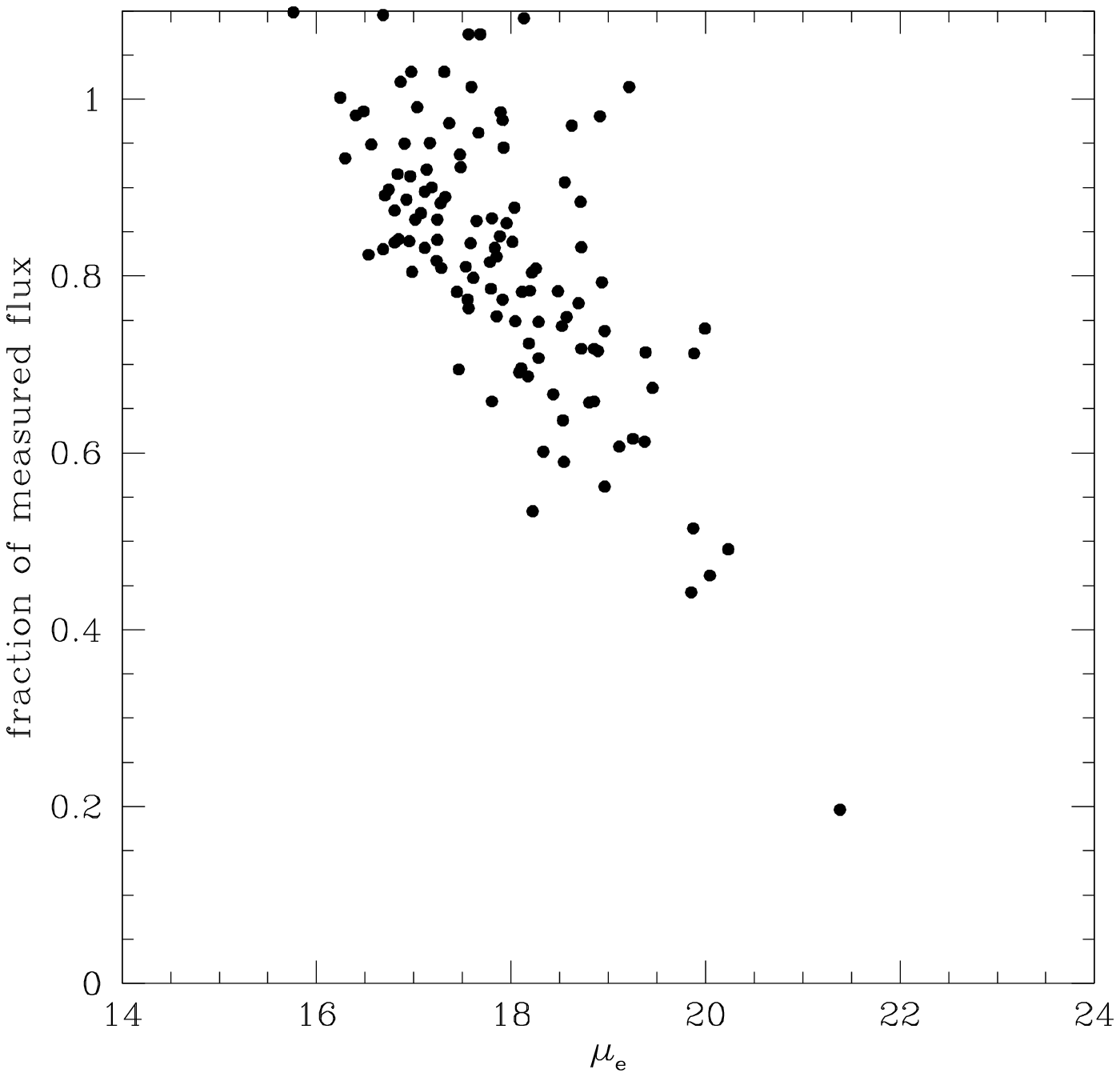,height=8truecm}
\psfig{figure=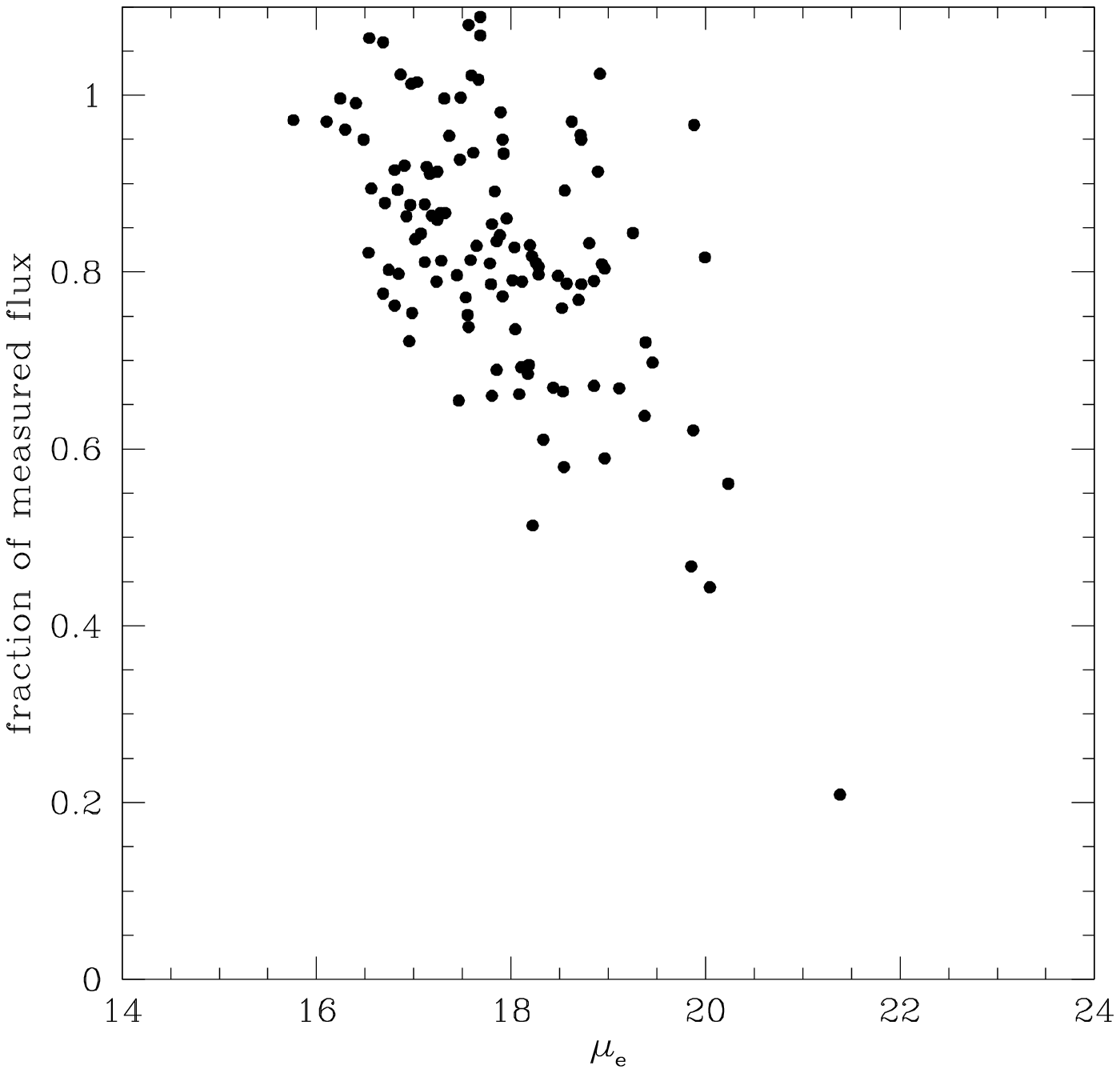,height=8truecm}}%
\caption[h]{Fraction of flux lost, vs isophotal
effective brightness, for galaxies both in the Pahre' and 2MASS samples. 
Left panel: isophotal 2MASS magnitude used by Kochanek et al., right panel:
hybrid Kron magnitude used by 2dF. A few outliers have brighter isophotal
and Kron
magnitudes than total one, possibly due to nearby objects not perfectly
handled by the 2MASS pipeline. 
}
\end{figure*}


\begin{figure*}
\psfig{figure=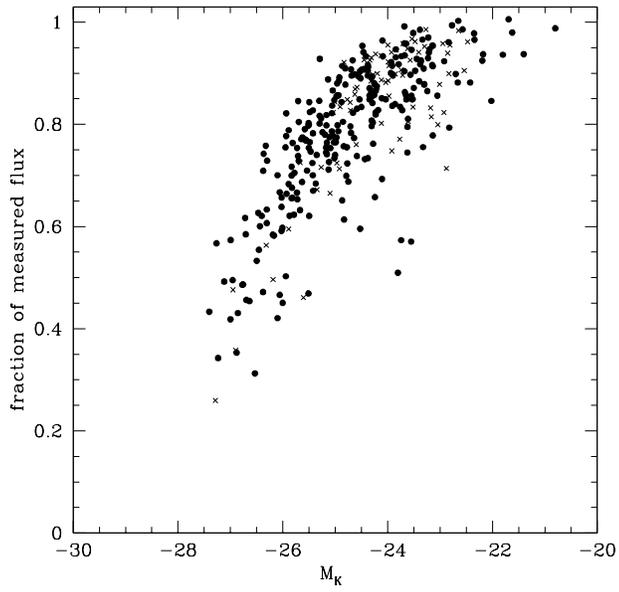,height=8truecm}%
\caption[h]{
Fraction of the flux inside the 3 arcsec aperture at $z=0.6$. Symbols as
in Fig 1. The same plot for galaxies at $z=0.3$ is qualitatively similar,
except for a larger missing flux, by 0.1, for all but the faintest
galaxies.}
\end{figure*}

\end{document}